\documentclass[12pt]{article}

\newcommand{\be}{\begin{equation}}
\newcommand{\ee}{\end{equation}}
\newcommand{\bea}{\begin{eqnarray}}
\newcommand{\eea}{\end{eqnarray}}
\newcommand{\nn}{\nonumber}
\newcommand{\bit}{\begin{itemize}}
\newcommand{\eit}{\end{itemize}}
\newcommand{\no}{\noindent}

\begin{document}

{\sf 

\title{The Geometry of M-Branes Wrapping Special Lagrangian Cycles}
\author{Ansar Fayyazuddin\footnote{email: Ansar\_ Fayyazuddin@baruch.cuny.edu} $^1$ 
and Tasneem Zehra Husain\footnote{email: tasneem@physics.harvard.edu} $^2$}

\maketitle

\begin{center}

{\it 
$^1$ Department of Physics, Baruch College, City University of New York\\ 

$^2$ Jefferson Physical Laboratory, Harvard University, Cambridge, MA 02138
}

\end{center}

\vspace{0.5cm}

\begin{abstract}
We characterize the geometry produced by M5-branes wrapping a Special Lagrangian 3-cycle in a Calabi-Yau 3-fold. The presence of the brane replaces the  
the Calabi-Yau by a real manifold with an almost complex structure. We show that, in this classification, a distinguished (1,1) form as well as a globally defined (3,0) form play an important role. The requirements of supersymmetry preservation impose constraints on these
structures which can be used to classify the background.  
\end{abstract}

\vspace{-16.5cm}
\begin{flushright}
HUTP-05/A0021 \\
BCCUNY-HEP /05-03 \\
hep-th/xxxxxxx
\end{flushright}

\thispagestyle{empty}

\newpage

\tableofcontents

\section{Introduction}
Characterizing supersymmetric string theory backgrounds is 
an important part of the string theory program.  Purely geometric 
backgrounds, with no other fields turned on, can be beautifully 
characterized and, as a result, completely classified.  Such backgrounds 
are Ricci flat geometries with reduced holonomy and 
additional intricate structures depending on their dimensions.  

The most well-studied of these purely geometric backgrounds are 
the Calabi-Yau manifolds in even dimensions.  Surprisingly, their 
discovery and characterization in terms of Hodge numbers, Kahler 
geometry, distinguished highest rank holomorphic forms etc has not 
been accompanied by an explicit construction of metrics on the compact 
versions of these manifolds. In fact, only a handful of metrics for 
non-compact Calabi-Yau manifolds are known.  Nevertheless, these manifolds 
have been crucial in our understanding of string theory in lower dimensions.  
The reason is that the existence of the rich structures that 
define the Calabi-Yau is enough to get a wealth of information; 
it is these structures that have been the source of our understanding of 
string theory compactification on Calabi-Yau manifolds.  

In a similar vein we can ask if it is possible to characterize and perhaps, more 
ambitiously, classify all supersymmetric string theory backgrounds. 
One subclass of problems in this more general scheme is to understand the 
geometry produced by branes wrapping cycles in Calabi-Yau manifolds.  We 
know that there are volume minimizing cycles in Calabi-Yau manifolds which 
sometimes yield supersymmetric configurations when branes are wrapped 
around them.  We also know that these wrapped branes will change 
the geometry drastically.  For instance, the Ricci flatness condition of
the Calabi-Yau will no longer be preserved.  Certain structures may survive, however.  As an example, if we were to wrap a brane on a holomorphic cycle in a Calabi-Yau, we would expect the resulting geometry to still have a complex structure compatible with that of the underlying Calabi-Yau.  

In this paper we study branes wrapping Special Lagrangian Cycles (SpelLs) in 
Calabi-Yau 3-folds in the context of M-theory.  SpelLs pose challenges of a 
new order compared to those of holomorphic cycles.  While holomorphic cycles 
preserve the complex structure of the underlying manifold, SpelLs are 
inherently real sub-manifolds and seem to destroy, at first sight, much of 
what makes Calabi-Yau manifolds tractable.  In this paper we will give a 
characterization of the full geometry of branes wrapping SpelLs.  

This paper is organized as follows: In Section 2 we define Special Lagrangian
cycles of Calabi-Yau manifolds and, using a brane probe, compute the Killing Spinors of an M5-brane wrapping a Special Lagrangian 3-cycle. In Section 3, we argue, the Killing Spinors found in the previous section play the role of the variation parameter in the full supergravity background and that this condition implies the existence of an almost complex structure.  Using the isometries of the brane configuration we give an ansatz for the metric
and four-form field strength. Finally, we study how the preservation of supersymmetry constrains these ansatze for the metric and field strength. Once
the constraints have been solved, we impose the Bianchi Identity and equations
of motion for F, thus guaranteeing that we have found a (bosonic) supersymmetric solution to the equations of motion of 11-d supergravity. Since some of these
constraints might seem a bit abstract, we work out the case of a planar M5-brane in Section 4. We end in Section 5 with conclusions. The Gamma matrix representations we used are written out explicitly in an Appendix.

\section{Special Lagrangian Cycles and their Killing Spinors}

Calabi-Yau manifolds come equipped with a Kahler form $\omega$ which is a 
distinguished member of $H^{(1,1)}$ as well as a unique holomorphic 
(n,0)-form $\Omega$.  A Special Lagrangian sub-manifold, $\Sigma$, 
of a Calabi-Yau n-fold is defined by the following set of conditions:
\begin{eqnarray}
\omega |_{\Sigma} &=& 0 \nonumber \\
\Re (e^{i\theta_\Sigma} \Omega) |_{\Sigma} &=& d\mbox{vol}(\Sigma)
\end{eqnarray}
where $\theta_{\Sigma}$ is a $\Sigma$-dependent constant phase.  
In other words, the pullback of the Kahler form vanishes on $\Sigma$ and, 
up to a phase, the pullback of $\Omega$ is the volume form on $\Sigma$.  
In fact, $\Omega$ 
gives a BPS bound on n-cycles so that only ones which saturate this bound 
are minimum volume cycles in their homology class. $\Omega$, thus, 
is a calibrating form.  

Given a SpelL $\Sigma$ we would like to answer the question: what
fraction of supersymmetry is preserved by an M5-brane wrapping it.  
The Killing spinors of a $p$-brane with embedding coordinates $X^{A}$ 
satisfy the following projection condition \cite{BBS}:
\begin{equation}
\epsilon = \frac{1}{p!} \epsilon^{\alpha_0\alpha_1 ... \alpha_p}
\Gamma_{A_0A_1 ... A_p} \partial_{\alpha_0} X^{A_0}
\partial_{\alpha_1} X^{A_1} ....
\partial_{\alpha_p} X^{A_p} \epsilon
\label{KSeqn}
\end{equation}
where $\epsilon$ is a Majorana spinor in 11 dimensions.  

Choosing static gauge along the directions $012$, the condition on the 
Killing spinors $\epsilon$ for an M5-brane wrapping a Special Lagrangian 
3-fold, is given by (\ref{KSeqn}): 
\begin{equation}
\epsilon^{abc}
\Gamma_{0 1 2} \Gamma_{i j k}
\partial_{a} X^{i} \partial_{b} X^{j}
\partial_{c} X^{k} \epsilon =
\epsilon
\label{dpsi}
\end{equation}
where ${\sigma}^a \dots {\sigma}^c$ are coordinates on the Special 
Lagrangian three-cycle and the $X^i$ are coordinates in the Calabi-Yau. 
Using complex coordinates on the Calabi-Yau, the projection condition becomes:
\bea
&&\frac{1}{2}\epsilon^{abc}
\Gamma_{0 1 2}[ \Gamma_{MNP}
\partial_{a} Z^M \partial_{b} Z^N 
\partial_{c} Z^P +  3\Gamma_{{M}\bar{N}\bar{P}}
\partial_{a} Z^{M} \partial_{b} Z^{\bar N}\partial_{c} Z^{\bar P} \nonumber \\
&+& 3 \Gamma_{\bar{M}\bar{N}{P}}
\partial_{a} Z^{\bar M} \partial_{b} Z^{\bar N}
\partial_{c} Z^{P}
 + \Gamma_{\bar{M}\bar{N}\bar{P}}
\partial_{a} Z^{\bar M} \partial_{b} Z^{\bar N}
\partial_{c} Z^{\bar P}]\epsilon =
\epsilon
\eea  
Since a Calabi-Yau is a complex manifold there is a choice of frame 
reflecting the complex structure - so that the only non-zero frame vectors have 
the form $e^a_A$ where $a, A$ are either both holomorphic or both 
anti-holomorphic indices.  The frame is defined as usual so that the metric 
on the Calabi-Yau $G_{M\bar{N}} = \eta_{m\bar n}e^m_Me^{\bar n}_{\bar{N}}$.  

If we define $\gamma_m = e_m^M\Gamma_M$ for $m=1,2,3$, these gamma matrices 
satisfy the flat space Clifford algebra which is identical, as is well known, 
to the creation annihilation algebra for 3 families of fermions.  
Thus a spinor can be represented by a set of states in a Fock space.  
We can define the vacuum state to be $ \epsilon_{000}$ and declare 
$\gamma_m$ to be annihilation operators, so that $\gamma_m \epsilon_{000} =0$. The remaining states can then be labeled by their occupation numbers corresponding to the action of the gamma matrices. We will use this construction to express the Killing spinor as a linear combination of Fock space states.

To avoid confusion we introduce $U, V, W$ as holomorphic coordinates on the Calabi-Yau.  Since the pullback of the (3,0)-form $\Omega$ is the 
volume form on the SpelL: 
$$\epsilon^{abc}\epsilon_{mnp}e^m_Ue^n_Ve^p_W\partial_{a} U \partial_{b} V
\partial_{c} W = 1$$ we can impose the condition (\ref{dpsi}) to find that generically \footnote{When compactifying on a Calabi-Yau 3-fold of SU(3) holonomy the only spinors that survive compactification are the SU(3) singlet spinors $\epsilon_{000}$ and $\epsilon_{111}$.  Similarly when compactifying on a Calabi-Yau 3-fold of smaller holonomy (for example $T^6$, $C^3$, $K3\times T^2$) the same spinors survive as long as the SpelL 3-cycle is non-trivially embedded in these manifolds.  If the SpelL is trivially embedded in these latter cases, then a greater amount of supersymmetry will be preserved.  We will study the consequences the minimum amount of symmetry being preserved, the other cases will be sub-cases. }:
\begin{equation}
\epsilon_{001} = \epsilon_{010} =  \epsilon_{001} 
=  \epsilon_{110} = \epsilon_{101} =  \epsilon_{110} = 0
\end{equation}
\no
The only components that survive are
$\epsilon_{000}$ and $\epsilon_{111}$, and these must obey
\bea
\gamma_{012}\gamma_{u v w} \epsilon_{111} 
&=& \epsilon_{000} \nonumber \\
\gamma_{012}\gamma_{{\bar u} {\bar v} {\bar w}} \epsilon_{000}
&=& \epsilon_{111}
\label{susyconst}
\eea

\no
It is convenient to express the flat gamma matrices in a direct product form.  This form is described in detail in the appendix.  The direct product structure exploits the split between the 3-complex dimensional space and the remaining 5-dimensional Minkowski space.  Using this split we can express 
the Killing spinor as follows
\bea
\epsilon &=& \epsilon_{000} + \epsilon_{111} \nonumber \\
&\equiv& \psi \otimes \eta_{000} + \chi \otimes \eta_{111}
\eea

\no
The Majorana condition on $\epsilon$ expressed in the gamma matrix basis 
of the appendix is $\gamma_{10}\epsilon = \epsilon^*$.   
Since $ \eta_{111} = \eta_{000}^*$ the Majorana condition implies:
\bea
\gamma_5 \psi &=& \chi^*  \;\;\;\;\;\;\;\;\;\;\;\;\;\;
\gamma_5 \psi^* = - \chi \nonumber \\
\tilde{\gamma_7} \eta_{000} &=& \eta_{000}  \;\;\;\;\;\;\;\;\;\;
\tilde{\gamma_7} \eta_{111} = - \eta_{111}
\eea
where $\psi*$ is the complex conjugate of $\psi$. Imposing now the
constraint (\ref{susyconst}) we find
\be
\gamma_{012}\psi = - \chi \;\;\;\;\;\;\; {\rm and} \;\;\;\;\;\; 
\gamma_{{\bar u} {\bar v} {\bar w}} \eta_{000} = \eta_{111} 
\ee

\no
These relations enable us to express the Killing spinor as: 
\be
\epsilon = \psi \otimes \eta_{000} +i\gamma_3 \gamma_5 \psi \otimes \eta_{111} 
\label{epsilon}
\ee
with the additional constraint:
\be
\psi^* = i \gamma_3 \psi 
\ee 
\no
The wrapped M5-brane then preserves $\frac{1}{8}$ of the spacetime
supersymmetry, corresponding to the 4 real degrees of freedom in $\psi$ 
which satisfy the
above conditions.  In what follows it will help to remember that $\psi, 
\gamma_3\psi, \gamma_5\psi, \gamma_3\gamma_5\psi$ are independent 
spinors as well as different Fock states built from $\eta_{000}$.

\section{Solving for the supergravity background for wrapped branes}
\label{ansatz}

\no
\subsection{Almost complex structure and the Killing spinor in the full background}
\no
In the previous section we found the fraction of supersymmetry preserved by branes wrapping SpelLs.  Our analysis was at the level of the "probe brane" approximation - we found the supersymmetry preserved by the brane without considering the modification of the geometry due to the presence of the brane.  We found that the supersymmetry preservation condition referred to gamma matrices with holomorphic and anti-holomorphic indices.  The notion of holomorphicity was inherited from the complex structure defined on the underlying Calabi-Yau.  

We expect that these supersymmetry preservation conditions should continue to hold true even when we take the full geometry into account.  This expectation is based on the fact that supersymmetry is preserved in discrete fractions and we wouldn't expect to be able to continuously change the amount of supersymmetry.  In the supergravity approximation to string theory we have classical branes which can be turned on continuously (i.e. the number of branes can be taken to be a continuum).  As we slowly turn on branes we expect that supersymmetry will stay the same fraction as it was originally in the probe approximation.  

We are thus led to ask how one can interpret the Killing spinor in terms of the structures of the full geometry?  It is actually easy to see that since the Killing spinor equation only refers to flat (or tangent space) holomorphic indices on the gamma matrices we only need an almost complex structure to state the Killing spinor equation - i.e. it is not necessary to have a complex structure. 

With this insight we will assume that the full geometry, while real, allows a global almost complex structure.  This is our key assumption which allows us to carry out the remainder of our program.

For completeness let us state the difference between having a complex structure and an almost complex structure in a more pedestrian way.  On a complex manifold we can classify tensors according to the number of holomorphic and anti-holomorphic differentials.  These differentials can be expressed as $dz^i$ and $dz^{\bar i}$ respectively, where $z^i$ are holomorphic coordinates on the manifold.  A tensor of type (p,q) has p holomorphic indices and q anti-holomorphic indices.  Almost complex manifolds, on the other hand, share with complex manifolds the property that they have a notion of (p,q) tensors {\it but} without the differentials being necessarily expressed in terms of holomorphic and anti-holomorphic coordinates on the manifold.  Thus a manifold may be an almost complex manifold without being a complex manifold.  

A convenient basis of (1,0) forms on an almost complex manifold of real dimension 2n can be given in terms of a distinguished set of frames.  In general we define a frame $e^a_I$ through the relation: $g_{IJ} = \delta_{ab}e^a_Ie^b_J$.  Any two frames are related by an SO(2n) rotation acting on the $a, b$ indices.  A complete basis of one-forms is given by $e^a=e^a_Idy^I$.  If there is an almost complex structure then it allows one to find a distinguished class of frames in which half of the $2n$ forms $e^a$ are (1,0)-forms while the other half are complex conjugates of the (1,0) forms.  One can label the (1,0) differentials with unbarred indices $e^m=e^m_Idy^I$ while the (0,1) forms are barred $e^{\bar m}=e^{\bar m}_Idy^I = (e^{m}_I)^*dy^I$.  

Any (p,q) form will be expressed as a sum over terms which are all products of p (1,0) forms and q (0,1) forms.  A r-form will in general be a sum over (p,q) forms such that p+q=r.  

If we want to preserve the almost complex structure (i.e. we don't want to mix (1,0) and (0,1) forms) we can only make U(1) $\times$ SU(n)$\subset$ SO(2n) rotations on the frame, so that the (1,0) forms transform in the fundamental representation of SU(n) with U(1) charge 1, while (0,1) forms transform in the anti-fundamental with U(1) charge -1.  The metric is given by the relation $$G_{IJ} = \eta_{m\bar{n}}(e^{m}_Ie^{\bar n}_J + e^{\bar n}_Ie^{m}_J)$$  Thus the metric and the almost complex structure define a basis of (1,0) forms up to a U(1)$\times$SU(n) rotation.  We can form another U(1)$\times$SU(n) invariant combination:
\be
B_{IJ} = \eta_{m\bar{n}}(e^{m}_Ie^{\bar n}_J - e^{\bar n}_Ie^{m}_J).
\ee  
which is a (1,1)-form in our classification.  $B$ is the analog of the Kahler form for almost complex manifolds and will be an important quantity for us.  

Finally, specializing to n=3, we can form the combination
\be
\Omega = e^u\wedge e^v\wedge e^w
\ee
which is a (3,0) form in our classification.  Although one can always form this combination, in general it will not be a globally defined 3-form.  We will be able to express physical quantities in terms of $\Omega$ and will thus establish that it is in fact a globally defined (3,0)-form.  It should also be clear that $\Omega$ is not invariant under the U(1) transformation defined above.  

\subsection{General method}

Having dispensed of the problem of interpreting the Killing spinor for the full geometry and, in the process, uncovering a rather elaborate apparatus involving almost complex structures, we move on to further elaborating the ingredients going into the construction of our background.  In the remainder of this section we will present a metric and 4-form ($F$) ansatz consistent with the isometries we require \cite{danda}.  

We then proceed to find the constraints on the metric and $F$ from requiring that the supersymmetry variation of the gravitino vanishes if the variation parameter is taken to be the Killing spinor (\ref{epsilon}).  Once we have solved the constraints on $F$ and the metric we will impose the Bianchi identity $d*F=0$ and the equation of motion $dF=J$ where $J$ is the source describing the location of the wrapped M5-brane.  Requiring all of this guarantees that we have a supersymmetric solution to the equations of motion.

\subsection{The Metric and Four-Form Ansatz}
To discuss the metric more concretely we first establish some conventions.  Before introducing the M5-brane we have M-theory on $R^{4,1}\times CY$.  Besides any isometries of the Calabi-Yau this space enjoys an SO(4,1) isometry in the $R^{4,1}$ factor.  We will take the $R^{4,1}$ factor to have coordinates $x^0, x^1, x^2, x^3, x^{10}$.  The wrapped M5-brane will span $x^0, x^1, x^2$ as well as a SpelL 3-cycle inside the Calabi-Yau.  At the very least, then, the brane will preserve a SO(2,1)$\times$SO(2)$\subset$SO(4,1), with SO(2) the group of rotations in $x^3,x^{10}$ plane.  We make these isometries explicit in the ansatz:
\be
ds^2  = {H_1}^2 \eta_{\mu \nu} dx^{\mu} dx^{\nu} + 
G_{I J} dy^{I} dy^{J} +
H_2^2 \delta_{\alpha \beta} dx^{\alpha} dx^{\beta}
\label{standard}
\ee
Where ${\mu},{\nu}$ run over $0,1,2$ and ${\alpha}, \beta$ run over ${3, 10}$.  The indices I and J run over $1,..,6$ and $y^I$ are coordinates over the part of space that was originally a Calabi-Yau before the presence of the branes transformed it.  In addition, we will assume that the metric preserves Poincare invariance in the $0,1,2$ directions, barring any explicit dependence on these coordinates.  If we move far away from the brane we should recover the original $R^{4,1}\times CY$ space; $G$ should approach the Calabi-Yau metric of the underlying manifold and $H_{1}, H_2 \rightarrow 1$.  

The 4-form field strength $F$ is defined through a 3-form potential $F=dA_3$.  Since M5-branes couple "magnetically" to the 3-form potential, the Bianchi identity and equation of motion for $F$ are interchanged.  $d*F + F\wedge F=0$ is assumed to be identically true while $dF = J$, where $J$ is the source current specifying the charge and location of the M5-brane.  The most naive form one can take for $F$ is such that $d*F=0$ identically so that the only non-zero components are $(*F)_{012IJKL}$ and $(*F)_{012IJK\alpha}$.  This would imply that one would only have to consider the following non-zero components for $F$: $F_{IJK\alpha}$ and $F_{IJ3(10)}$.  We take this ansatz in the remainder of the paper.  It has to satisfy the self-consistency check $F\wedge F=0$.  Notice that this is not the most general ansatz consistent with the isometries of the problem but there are good reasons to suspect that this ansatz is sufficient.

\subsection{The Supersymmetric Variation of the Gravitino}

Given the ansatz for the metric and 4-form $F$ we can ask what the constraints are within that ansatz if we are to preserve the supersymmetries described in the previous section.  To this end we take our metric and 4-form and plug them into the supersymmetry variation equation for the gravitino in 11-dimensional supergravity in a purely bosonic background:

\be
\delta_{\epsilon}\Psi_{I} = (\partial_{I}  + \frac{1}{4} \omega_I^{ij} 
\hat{\Gamma}_{ij}
+ \frac{1}{144}{\Gamma_{I}}^{JKLM}F_{JKLM}
-\frac{1}{18}\Gamma^{JKL}F_{IJKL})\epsilon = 0
\label{susy}
\ee
We will take $\epsilon$ to be given by (\ref{epsilon}).  
The requirement $\delta_{\epsilon} \Psi = 0$ can then be expressed as 
the vanishing of a 
combination of linearly independent Fock space states. The 
coefficient of each state must hence be set to zero identically, 
giving us a set of constraints on the metric and field strength. 

To summarize: the supergravity solution for an M5-brane wrapping a SpelL 3-cycle in a 
CY 3-fold is found by demanding that (\ref{susy}) is satisfied
for a metric of the form (\ref{standard}), when the variation parameter
$\epsilon$ is a Killing spinor of the form (\ref{epsilon}). This gives 
rise to a set of equations which enable us to constrain the geometry and 
find the components of the four-form field strength.\\

From (\ref{susy}) we find that the Killing spinor $\epsilon$ can be expressed as 
\be
\epsilon = H^{-1/12} \hat{\psi} \otimes \eta_{000} - \gamma_5 
H^{-1/12} i \gamma_3 \gamma_5 \hat{\psi} \otimes \eta_{111} 
\ee
where $\psi^* = i \gamma_3 \psi$ and $\hat{\psi}$ is a constant spinor.  
In addition the function H is related to $H_1$ and $H_2$ through
\be
H_1^2 = H_2^{-1} \equiv H^{-1/3}
\ee

\no
The field strength is found to be
\bea
F &=& \frac{i}{16}\epsilon_{\alpha\beta}\partial_\beta\ln H
[\Omega - \bar{\Omega}] \wedge dx^\alpha \\
&+& \frac{i}{16}H^{-1/3}\partial_L H 
[\Omega^{L}_{\;\;IJ} - \bar{\Omega}^{L}_{\;\;IJ}] dy^I\wedge dy^J\wedge 
dx^3\wedge dx^{10} \nonumber \\
&+&\frac{i}{16} [(P_-)_{KJ} \Omega_{LMN}
- (P_+)_{KJ} \bar{\Omega}_{LMN}] 
\epsilon_{\alpha\beta}\partial_{\beta
}G^{LJ}dy^{K}\wedge dy^M\wedge dy^N\wedge dx^\alpha \nonumber \\
&-&\frac{i}{8}H^{2/3} (P_-)_M^{\;I} [\Omega_{LJK}\partial_IB^{JK} 
+2\Omega_L^{JK}\partial_KG_{IJ}]dy^M\wedge dy^L\wedge dx^3\wedge dx^{10} \nn
\label{ff}
\eea
where $P_+$ and $P_-$ are projection operators defined as follows
\be
(P_+)_M^{\;\;\;N} = \frac{1}{2} (\delta_M^N + B_M^{\;\;\;N}) \;\;\;\;\;\;\;\;
(P_-)_M^{\;\;\;N} = \frac{1}{2} (\delta_M^N - B_M^{\;\;\;N})
\ee
Hence $P_+$ projects onto tensors of type (1,0) and $P_-$ onto those of type (0,1).\\

\no
In addition to giving an expression for the four-form field strength, supersymmetry preservation dictates a set of equations 
which can be solved to yield a number of independent constraints. 
\no
There are constraints on the $(3,0)$-form,
\be
\Omega^{IJK} \partial_{\alpha} \bar{\Omega}_{IJK} 
= 12 \partial_{\alpha} ln G
\ee
on the almost complex structure
\bea 
\partial_I [ H^{1/2} B^I_{\;\;J}] = 0 \\
d_6 [H^{-1/3} B \wedge B] = 0 
\eea
and then there are those that relate the two
\bea
\Omega \wedge d_6 B = 0 \\
\Omega_{IJK} \partial_{\alpha} B^{JK} = 0
\eea
Here $d_6 =dy^I\partial_I$ is the exterior derivative in the directions along the CY.

One can check from (\ref{ff}) that $F \wedge F$ vanishes identically.  
However, imposing the Bianchi identity $d*F=0$ leads to 
some interesting information. 
For one, we obtain a relation between the function H in our metric ansatz and the determinant of the 
metric on the 'Calabi-Yau' 
\be
Det \; G_{IJ} \equiv G = H
\ee
In addition we obtain the constraint
\be
\bar{\Omega} \wedge *_6 \; d_6\Omega =0
\ee
which tells us that $d_6\Omega$ which would be normally an element of $\Lambda^{(3,1)}\oplus\Lambda^{(2,2)}$ is actually just an element of 
$\Lambda^{(2,2)}$ alone. 
\no
Finally,
\be
d_6(\Omega - \bar{\Omega}) =0
\label{imomegaclosed}
\ee 
This is essentially the statement that $Im \; \Omega$, the form which is 'orthogonal'\footnote
{$Im \; \Omega \wedge Re \; \Omega = \Omega \wedge {\bar \Omega} = Vol \; form$} to the calibrating form $Re \; \Omega$ of
the SpelL, is closed. In hindsight, this is something we should have expected. In our earlier experience of 
exploring geometries that arise when M-branes wrap holomorphic cycles in Calabi-Yau manifolds, we found that once
the back-reaction of the brane was taken into account, the manifold did not necessarily remain Kahler; 
preservation of supersymmetry imposes a constraint to the effect that the form orthogonal to the calibration 
must be closed\footnote {For holomorphic 2-cycles, the calibration is 
simply the (would-be) Kahler form $\omega$. 
The above mentioned constraint can then be stated as follows $d * \omega = 0$ \cite{Tasneem}. For a 
holomorphic 4-cycle, the calibrating form is $\omega \wedge \omega$ and the corresponding constraint becomes 
$d * (\omega \wedge \omega) = 0$}. The constraint (\ref{imomegaclosed}) is the analogue of those earlier
constraints extended to this new situation where branes wrap SpelLs instead of holomorphic curves. 

If all of the above holds, we find we can express $*F$ as the exterior derivative acting on a six-form gauge potential
which is, as expected, the generalised calibration corresponding to an M5-brane wrapping a SpelL3-cycle.  
\be
*F  =  d [ \frac{1}{4} H^{-1/2} \; dx^{0} \wedge dx^{1} \wedge dx^{2} \wedge
(\Omega + \bar{\Omega})]
\label{*F}
\ee

\no
Using the constraints listed above, we can re-write the four-form (\ref{ff}) as 

$$ F = \frac{1}{4} H^{1/6} *_6 d_6[H^{1/2}( \Omega + \bar{\Omega})] \wedge dx^3 \wedge 
dx^{10} - \frac{i}{4} H^{-1/2} \epsilon_{\alpha \beta} \partial_{\beta} 
[H^{1/2} (\Omega -\bar{\Omega})] \wedge dx^{\alpha}$$
The explicit form of the metric is obtained upon solving the differential equation
\be
dF = J
\ee
where J is the source for the M5-brane. Even in the absence of a specific metric however, we have been able to 
uncover an extremely rich geometrical structure and use it to characterise the 
supergravity background of an M5-brane wrapping a Special Lagrangian 3-cycle.

\section{Illustrating the Geometry: The case of a flat M5-brane}
Consider flat space $R^{10,1}$ and write it as $R^{5,1}\times C^3$.  We define a complex structure on $C^3$ so that holomorphic coordinates are given in terms of the cartesian coordinates on $R^{10,1}$ as $U = x^4 + i x^5, V = x^6 + i x^7$, $W = x^8 + i x^9$.  

Now take an ordinary flat M5-brane with worldvolume $012468$.  The M5-brane is oriented along the $012$ directions as well as a "3-cycle" which is an $R^3$ (with coordinates $x^4, x^6, x^8$) in $C^3$.  It is easy to see that this is a Special Lagrangian cycle: the (3,0) form $\tilde{\Omega} = dU\wedge dV\wedge dW$, pulled back to $R^3$ gives the volume form and the Kahler form for the flat metric on $C^3$ vanishes on $R^3$.

\no
We know the full metric for an M5-brane in flat space, it is
\bea
ds^2 = H^{-1/3} (-dx_0^2 + dx_1^2 + dx_2^2 + dx_4^2 + dx_6^2 + dx_8^2) 
\nonumber \\
+ H^{2/3} (dx_5^2 + dx_7^2 + dx_9^2 + dx_3^2 + dx_{10}^2)
\eea
\no
where $H$ is a harmonic function with respect to the flat Laplacian in the transverse coordinates. The (1,1)-form associated with the metric\footnote{This is 
generally denoted by $\omega$ but to be consistent with the 
notation elsewhere in this paper, we stick to B.} is given by 
\bea
B &=& e^u \wedge e^{\bar u} + 
e^v \wedge e^{\bar v} + e^w \wedge e^{\bar w} \nonumber \\
&=& H^{1/6}(dU\wedge d\bar{U} + dV\wedge d\bar{V} + dW\wedge d\bar{W})
\eea
where small letters denote tangent space indices. 
The holomorphic three-form can be expressed as
$$ \Omega = e^u \wedge e^v \wedge e^w $$
where 
\bea
e^u &=& e^{u}_{U} \; dU + e^{u}_{\bar U} \; d{\bar U} \nn \\
&=&  \frac {1}{2} H^{-1/6} \; [ (1 + {\sqrt H}) dU + 
(1 - {\sqrt H}) d{\bar U}]
\eea
and similar expressions hold for $e^v$ and $e^w$.

It is now a simple exercise to show that $B$ and $\Omega$ satisfy all the constraints of the previous section. It is also easy to see explictly that the almost complex structure is non-integrable: $d_6e^m$ has a non-zero $(0,2)$ component.  

In order to reproduce completely our earlier results, we need only to 
show that the six-form gauge potential to which the M5-brane couples 
'electrically' is given by (\ref{*F})
$$A = \frac{1}{4} H^{-1/2} \; dx^{0} \wedge dx^{1} \wedge dx^{2} \wedge
(\Omega + \bar{\Omega})$$
This too, is simply done. Using the expressions for $e^u, e^v$ etc, 
$\Omega$ and $\bar{\Omega}$ can be written out explicitly
in terms of dU, dV \dots. We then find that
\be
A = \frac{1}{2} H^{-1} \; dx^0 \wedge dx^1 \wedge dx^2 \wedge 
dx^4 \wedge dx^6 \wedge dx^8 
\ee
which is the well-known expression for the gauge potential coupling 
to a flat M5-brane!  

The field strength, in flat space coordinates is 
\bea
F &=& \frac{i}{8} H^{-4/3} \partial_{10}H (e^u - e^{\bar u})
\wedge(e^v - e^{\bar v}) \wedge (e^w - e^{\bar w}) \wedge e^3 \nn \\
&-& \frac{i}{8} H^{-4/3} \partial_{3}H 
(e^u - e^{\bar u}) \wedge(e^v - e^{\bar v}) 
\wedge(e^w - e^{\bar w})\wedge e^{10} \nn \\
&-&\frac{i}{4} H^{-4/3}(\partial_{u}-\partial_{\bar u})H (e^v - e^{\bar
v})\wedge(e^w - e^{\bar w})\wedge e^3\wedge e^{10} \nn \\
&-& \frac{i}{4} H^{-4/3}(\partial_{v}-\partial_{\bar v})H (e^w - e^{\bar
w})\wedge(e^u - e^{\bar u})\wedge e^3\wedge e^{10} \nn \\
&-& \frac{i}{4} H^{-4/3}(\partial_{w}-\partial_{\bar w})H (e^u - e^{\bar
u})\wedge(e^v - e^{\bar v})\wedge e^3\wedge e^{10}
\eea

\section{Conclusions}

In this paper we studied M5-branes wrapping Special Lagrangian 3-cycles in Calabi-Yau 3-folds.  Our approach was based on finding the correct fraction of supersymmetry preserved.  We found that this led to an intricate geometry involving almost complex structures.  We showed how that geometry was highly constrained and that all the fields in the supergravity solution could be expressed in terms of the metric, the (1,1) form $B$ and the (3,0) form $\Omega$ which 
characterise the geometry.

Branes wrapping Special Lagrangian cycles have of course been studied before. 
The work most closely related to our own is that of Martelli and Sparks \cite{m&s} who studied wrapped branes on SpelLs using the language of G-structures\footnote{A G-structure is basically a group of locally defined forms which are invariant under the group G. These forms are typically constructed as bilinears of Killing spinors (see \cite{m&s} for a more comprehensive account)}. 
They argue that if an M5-brane wraps a SpelL 3-cycle
in a Calabi-Yau three-fold, then the back-reaction modifies the manifold such
that it no longer has SU(3) holonomy, but nevertheless an SU(3) structure remains. They then go on to elaborate that an SU(3) structure is 
specified by two vectors, a two-form and a three-form.
Though we do not use the language of G-structures explicitly, 
we have constructed these quantities  - they are the vectors $e^3$ and $e^{10}$, 
the two-form $B$ and the (3,0) form $\Omega$.  In the present work we go beyond the constructions of \cite{m&s} in specifying the geometry more completely.  We also hope that our construction makes more intuitive the origins of this geometry. 

In \cite{memam}, M5-branes wrapping SpelL 3-cycles in Calabi-Yau threefolds
are studied as 2 dimensional branes in a d=5 supergravity theory.  In this approach one can only study the geometry transverse to the Calabi-Yau but not the modification of the geometry of the Calabi-Yau itself. 

Branes wrapping SpelLs are studied in \cite{gauntlett}, within the context 
of $d=7$ gauged supergravity, with gauge group $SO(4)$ or $SO(5)$ 
which is the consistent truncation of a higher dimensional supergravity theory reduced on a sphere of appropriate dimension. The solutions are constructed in seven dimensions and then lifted back up to ten (or eleven) dimensions. 
One of the limitations of this approach is that one necessarily has to assume  the isometries of a round $S^4$ (or $S^3$). Another factor which distinguishes their analysis from ours is that they take the near horizon limit and find the 
supergravity solution there. Thus, only the local geometry of the calibrated
cycle enters into the construction. 

There are a number of interesting directions one can take from here.  One such is to apply these considerations to Calabi-Yau 2- and 4-folds where the existence of a continuous class of complex structures allows one to smoothly go between SpelLs and holomorphic cycles. In these cases we should be able to connect our approach here to previous work on holomorphic cycles. We are currently investigating these ideas.  Another set of problems, of course, involves finding new supergravity solutions satisfying all the constraints elaborated in this 
paper.\\

\no
{\Large \bf Acknowledgements}\\

We are grateful to Dileep Jatkar for collaboration during the early stages of this work. AF would like to thank Ingemar Bengtsson for useful discussions.  AF would like to acknowledge funds from the Swedish Vetenskapsr{\aa}det (VR) during the initial stages of the research presented here.  AF would also like to acknowledge the hospitality of Tufts University, Harvard University,  Stockholm University, and Uppsala University at various stages of this project.  TZH would like to acknowledge funding from VR.

\newpage
\no
{\Large \bf Appendix}\\

\no
\underline{\sf A: Representation of Gamma Matrices}\\

\no
A 4-dimensional real representation of Gamma matrices can be constructed out of Pauli matrices as follows:
\bea
\gamma_0 &=& i \; \sigma_1 \otimes \sigma_2 \nonumber \\
\gamma_1 &=&  1 \otimes \sigma_3 \nonumber \\
\gamma_2 &=& \sigma_2 \otimes \sigma_2 \nonumber \\
\gamma_3 &=& 1 \otimes \sigma_1 
\eea
\no
The 4-dimensional chirality operator is thus given by
\be
\gamma_5 = - i \gamma_0 \gamma_1 \gamma_2 \gamma_3 
= - \sigma_3 \otimes \sigma_2
\ee

\no
Similarly, a 6 dimensional imaginary representation of gamma matrices can be constructed.  For example:
\bea
\tilde{\gamma_1} &=& \sigma_2 \otimes 1 \otimes 1 \nonumber \\
\tilde{\gamma_2} &=& \sigma_1 \otimes \sigma_2 \otimes 1 \nonumber \\
\tilde{\gamma_3} &=& \sigma_1 \otimes \sigma_1 \otimes \sigma_2 \nonumber \\
\tilde{\gamma_4} &=& \sigma_1 \otimes \sigma_3 \otimes \sigma_2 \nonumber \\
\tilde{\gamma_5} &=& \sigma_3 \otimes 1 \otimes \sigma_2 \nonumber \\
\tilde{\gamma_6} &=& \sigma_3 \otimes \sigma_2 \otimes \sigma_3 
\eea
\no
It also proves useful to define Gamma matrices for the complex coordinates
\bea
\gamma_u = \frac{1}{2} \left[ \tilde{\gamma_1} + i \tilde{\gamma_2} \right] 
\nonumber \\
\gamma_v = \frac{1}{2} \left[ \tilde{\gamma_3} + i \tilde{\gamma_4} \right] 
\nonumber \\
\gamma_w = \frac{1}{2} \left[ \tilde{\gamma_5} + i \tilde{\gamma_6} \right]
\eea
\no
The chirality operator in these 6-dimensions is 
\bea
\tilde{\gamma_7} &=& i \tilde{\gamma_1} \tilde{\gamma_2} \dots 
\tilde{\gamma_6}  =  8 \gamma_{u {\bar u}}\gamma_{v {\bar v}}\gamma_{w {\bar w}}
\nonumber \\ 
&=& \sigma_3 \otimes \sigma_2 \otimes \sigma_1 
\eea

\no
We can now write down the 11-dimensional $32 \times 32$ Gamma matrices explicitly:
\bea
\Gamma_{\mu} &=& \gamma_{\mu} \otimes \tilde{\gamma_7} \nonumber \\
\Gamma_{i+3} &=& 1 \otimes \tilde{\gamma_i} \\
\Gamma_{10} &=& \gamma_5\otimes\tilde{\gamma}_7 \nonumber
\eea
where $\mu = 0 \dots 3$ and $i = 1 \dots 6$. 
All the gamma matrices are imaginary with the exception of $\Gamma_{10}$ which is real.  
\no
Thus the Majorana condition is $\epsilon^* = \Gamma_{10}\epsilon$.

}
\end{document}